\documentclass[twocolumn,preprintnumbers,aps,amsmath,amssymb,superscriptaddress,a4paper]{revtex4}
\usepackage{verbatim}

\pdfoutput=1

\usepackage{amsmath}
\usepackage{amsthm}
\usepackage{amssymb}

\usepackage{bm}
\usepackage{bbm}
\usepackage{dsfont}

\usepackage{enumitem}

\usepackage{graphicx}
\usepackage{epstopdf}
\usepackage{tikz}
\usetikzlibrary{trees,arrows,calc}

\tikzstyle{level 1}=[level distance=1.875cm, sibling distance=1.5cm]
\tikzstyle{level 2}=[level distance=1.875cm, sibling distance=1.5cm]
\tikzstyle{level 3}=[level distance=1.875cm, sibling distance=1.5cm]
\tikzstyle{level 4}=[level distance=1.25cm, sibling distance=1cm]

\tikzstyle{bag} = [text width=4em, text centered]
\tikzstyle{end} = [->, minimum width=3pt]

\newcommand{\ket}[1]{|#1\rangle}

\newcommand{\expt}[1]{\langle #1 \rangle}

\theoremstyle{plain}

\theoremstyle{definition}

\begin{document} 

\setlength{\tabcolsep}{1ex}

\title{Organic molecule fluorescence as an experimental test-bed for quantum jumps in thermodynamics}

\author{Cormac Browne\footnote{Corresponding author e-mail: cormac.browne@physics.ox.ac.uk}}
\affiliation{Clarendon~Laboratory, University~of~Oxford, Parks~Road, Oxford, OX1 3PU, UK}

\author{Tristan Farrow}
\affiliation{Clarendon~Laboratory, University~of~Oxford, Parks~Road, Oxford, OX1 3PU, UK}
\affiliation{Center for Quantum Technologies, National University of Singapore, Republic of Singapore}

\author{Oscar C. O. Dahlsten}
\affiliation{Clarendon~Laboratory, University~of~Oxford, Parks~Road, Oxford, OX1 3PU, UK}
\affiliation{London~Institute, 35a South Street~Mayfair, London, UK}

\author{Vlatko Vedral}
\affiliation{Clarendon~Laboratory, University~of~Oxford, Parks~Road, Oxford, OX1 3PU, UK}
\affiliation{Center for Quantum Technologies, National University of Singapore, Republic of Singapore}
\affiliation{Department~of~Physics, National~University~of~Singapore, 2~Science~Drive~3, Singapore~117542}
\affiliation{Center~for Quantum~Information, IIIS, Tsinghua~University, Beijing, 100084, China}

\date{\today}

\begin{abstract}
We demonstrate with an experiment how molecules are a natural test-bed for probing fundamental quantum thermodynamics. Single-molecule spectroscopy has undergone transformative change in the past decade with the advent of techniques permitting individual molecules to be distinguished and probed. By considering the time-resolved emission spectrum of organic molecules as arising from quantum jumps between states, we demonstrate that the quantum Jarzynski equality is satisfied in this set-up. This relates the heat dissipated into the environment to the free energy difference between the initial and final state. We demonstrate also how utilizing the quantum Jarzynski equality allows for the detection of energy shifts within a molecule, beyond the relative shift.
\end{abstract}

\maketitle



Marrying the language of thermodynamics with quantum phenomena is giving rise to a quantum thermodynamics whose development defines a new frontier where the transfer of energy at the level of individual quantum objects can now be studied. The success of thermodynamics owes much to being open to experimental testing in a variety of systems \cite{Goold15}. These have included systems operating in the quantum regime and exhibiting features like quantum coherence, such as quantum dots~\cite{Hekking13}. Here we seek to demonstrate that the fluorescence of organic molecules forms a natural testbed for these theories. Our result provides an important proof of concept for the validity of quantum thermodynamics in this regime, and opens the door for more advanced tests of the theory.\\

Non-equilibrium  thermodynamics is evolving rapidly, and a key result is the Jarzynski equality \cite{Jarzynski97, Crooks99, Goold15}. This concerns the probability distribution of work into a driven system in contact with a heat bath. It essentially equates the average of the exponential work with something which is constant regardless of the driving rate. That constant is the exponential of the equilibrium free energy, i.e. the free energy between a thermal state at the initial boundary conditions and at the final respectively. A key use of this equation is to determine the equilibrium free energy difference from non-equilibrium experiments \cite{Fox03, Park03, Ytreberg04}. The equation also holds for quantum systems \cite{Jarzynski04, Quan08}. One important area within this field, which has received a large amount of recent attention, is the development of experimental techniques and protocols which can test the theoretical predictions \cite{Batalhao14, An15}.\\

Parallel developments in physical chemistry with the advent of single molecule spectroscopy~\cite{nobel15} over the past decade have opened up unprecedented opportunities for the study of single quantum systems. Unlike studies in bulk where spectroscopic signatures are washed out by the averaging effect of ensembles, these novel techniques now allow individual molecules to be identified, tracked and probed. Thus it is becoming possible to study energy transfers at the level of single molecules, some of which offer ideal test-beds owing to their well-defined spectroscopic signatures and quantum state dynamics. Their properties are extremely reproducible, more so than rival quantum systems such as quantum dots and nanocrystals~\cite{nobel15}.
\begin{figure}[h!]
\begin{centering}
\includegraphics[width=\columnwidth]{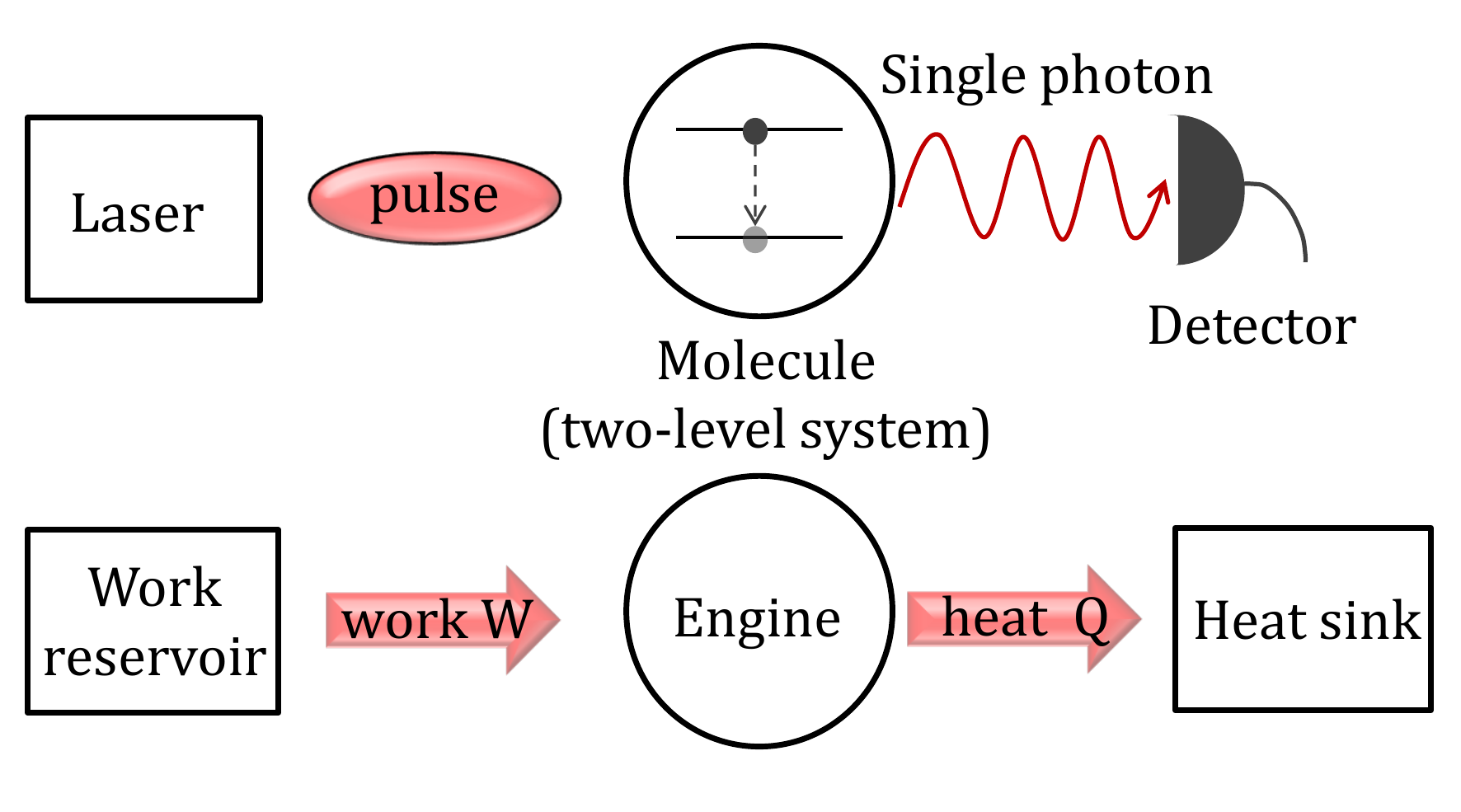}
\caption{\label{fig:engine} Schematic of a two-level molecule driven by laser pulses generating single photons from decaying electronic states with our thermodynamical representation of the process below it.}
\end{centering}
\end{figure}

\begin{figure}[h!]
\begin{centering}
\includegraphics[width=\columnwidth]{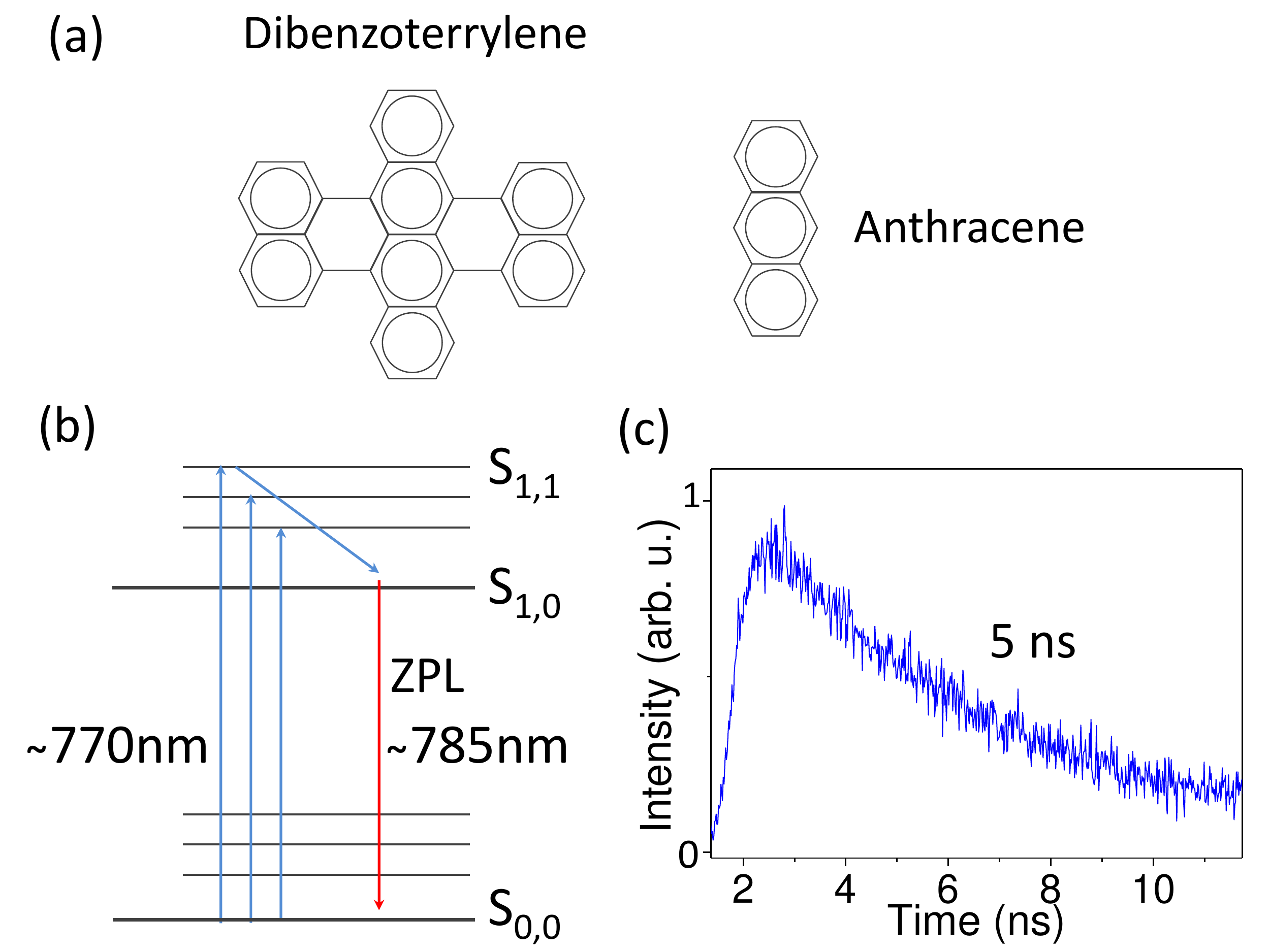}
\caption{\textbf{(a)} Dibenzoletrrylene (DBT) and anthracene structures consist of corrugated planar assemblies of aromatic hydrocarbons. Two possible DBT insertion sites in the extended anthracene host crystal lattice act as defects giving rise to two distinct spectral lines at 785~nm and 795~nm. In this study we chose to probe the brighter blue-shifted site with a characteristic transition line at 785~nm. \textbf{(b)} The energy level structure of DBT reduces to two-level atom-like states  typical of organic fluorescent dyes despite the relative complexity of the organic molecules. We pumped DBT above resonance at 4~K to generate excited states that decayed non-radiatively on pico-second timescales into the first excited state $S_{1,0}$. Decay from $S_{1,0}$ into the ground state $S_{0,0}$ gave rise to a sharp lifetime limited line (30Mhz), which, in a single molecule, gives rise to the emission of a single photon for each decay. The narrow width of the Zero Phonon Line (ZPL is lifetime-limited with a width of approx. 30MHz. \textbf{(c)} Waveform of the decay of excited electronic states in DBT at a temperature of 4K obtained by time-resolved fluorescence spectroscopy. The waveform represents a probability distribution of the arrival times of single photons from the decaying state rather than the shape of an optical signal. The time-resolved technique relies on time-correlated single photon counting of pulses obtained by exciting the sample with a 730~nm pulsed laser with a repetition rate of 80MHz. The probability of detecting more than one photon per excitation period is negligible. The excited state which gives to the ZPL emission line is found to have a characteristic lifetime of 5ns aftering fitting the exponential decay constant.}
\label{fig:DBT}
\end{centering}
\end{figure}

In this work we connect these two approaches, namely quantum thermodynamics and molecular spectroscopy. We consider the spontaneous emission of photons from a single excited organic molecule as a thermodynamic process. We determine the probability distribution of work and heat in the experiment, see~\ref{fig:engine}, and find that it satisfies the quantum Jarzynski equality. A key point is to treat the free space around the molecule as an extremely low excitation heat bath, and the fluorescent light as heat transferred into that environment. The laser applied initially corresponds to the driving force, as depicted in Fig.~\ref{fig:engine}. In the theoretical analysis we use quantum jumps. The core concept in the quantum jump formalism is that the system always occupies a distinct state that undergoes a stochastic evolution which enables it to transition between different energy levels \cite{Plenio98, Hekking13} as illustrated in Figs.~\ref{fig:DBT} and~\ref{fig:trajplot}. Furthermore, we demonstrate how use of the quantum Jarzynski equality can be used to infer absolute energy-level shifts in molecules from the emitted photons, as opposed to just the relative shift. We thus establish that these organic molecules are a natural and powerful test-bed for non-equilibrium thermodynamics and suggest how they have use in the enhanced detection of force fields.

\section{Results}

\begin{figure}[h!]
\begin{centering}
\includegraphics[width=5cm]{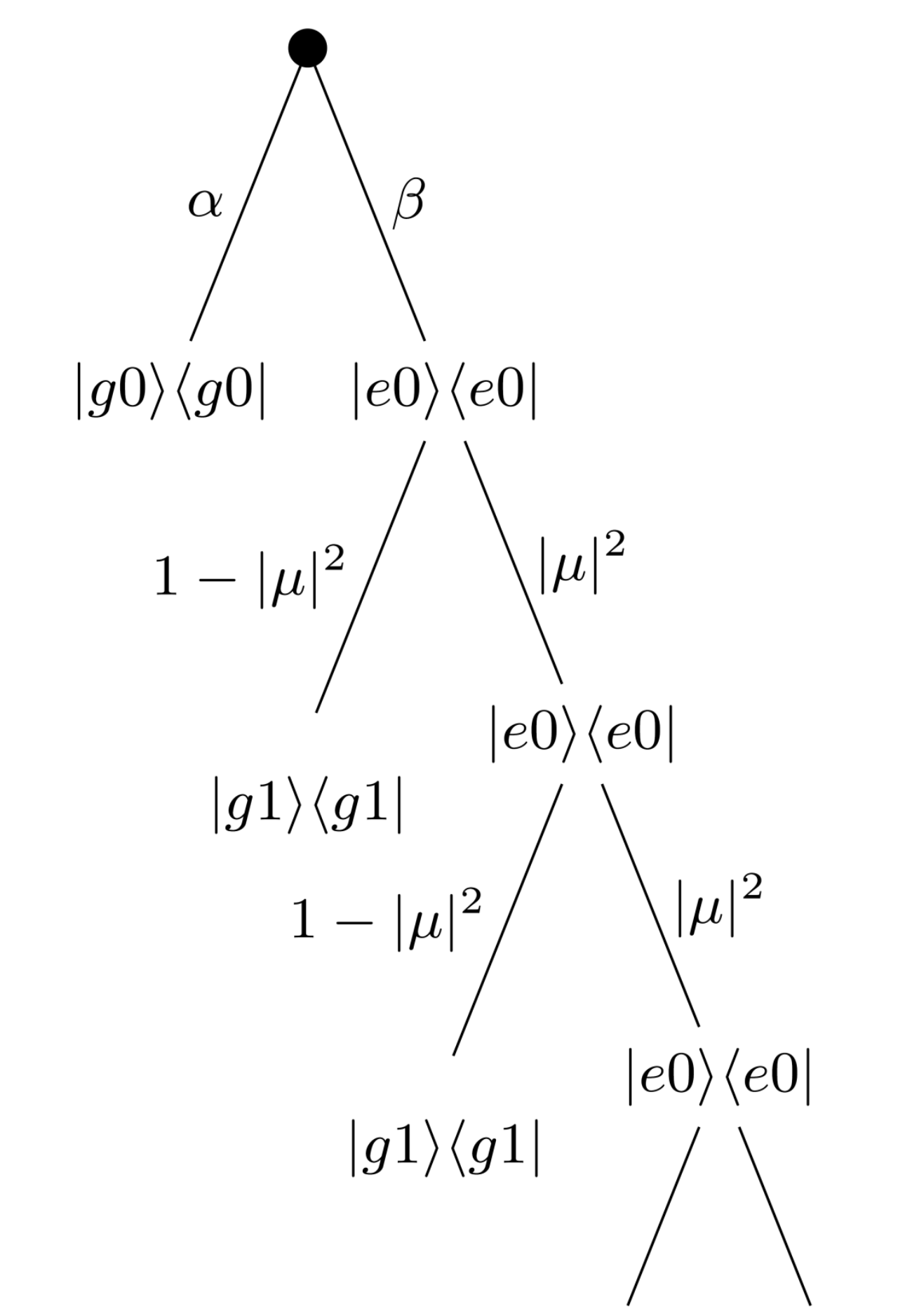}
\caption{A sample of the different trajectories the system can--according to the model-- take after interacting with the environment several times, after having been initialized in a mixed state. Each trajectory is formed by alternating a period of evolution under the joint Hamiltonian and then performing a projective measurement on the environment. As there is only one photon during the protocol and the probability of spontaneous excitation of the molecule is negligible, once the photon has moved into the environment there are no further branchings.}
\label{fig:trajplot}
\end{centering}
\end{figure}

\begin{figure}[h!]
\begin{centering}
\includegraphics[width=\columnwidth]{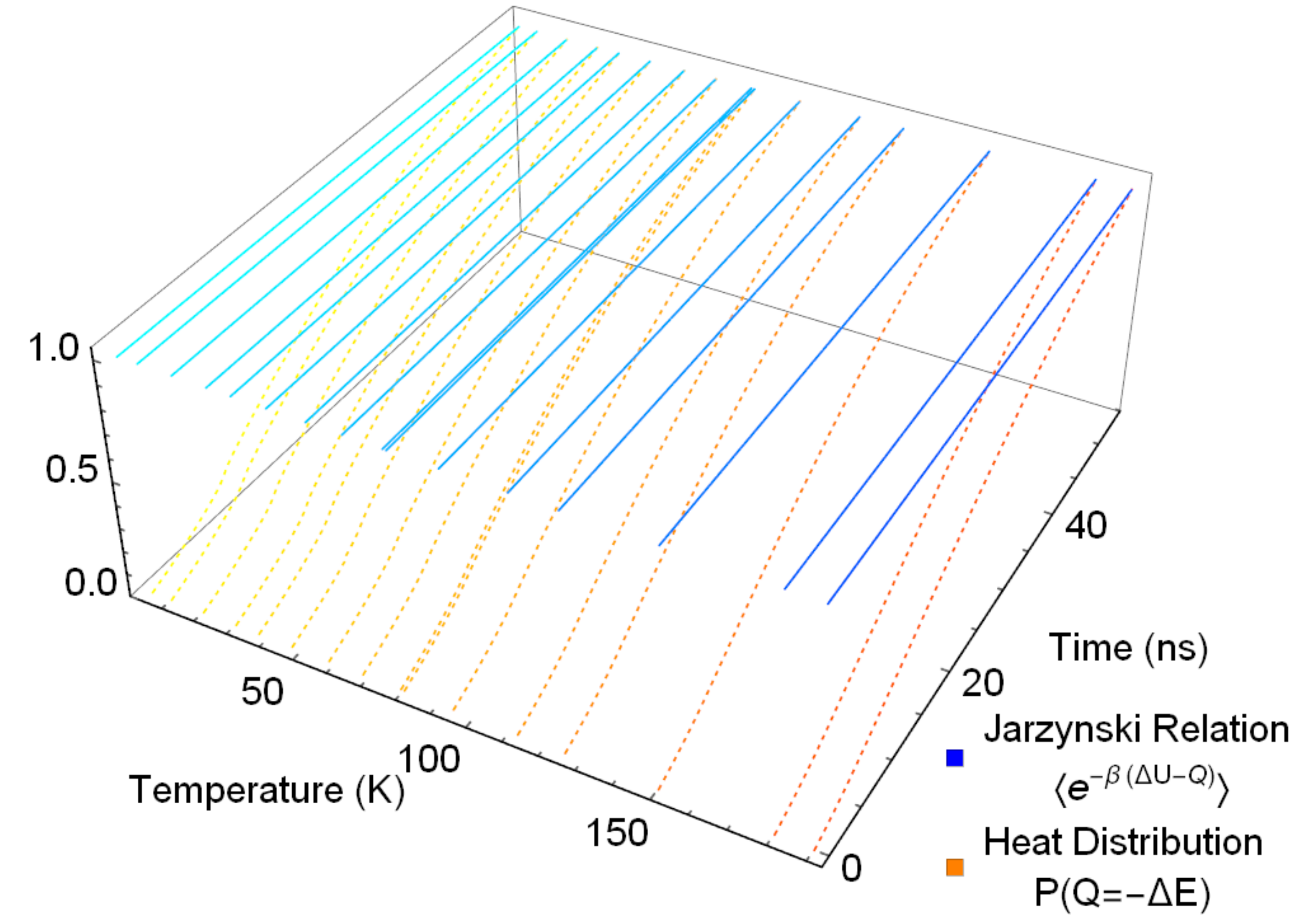}
\includegraphics[width=\columnwidth]{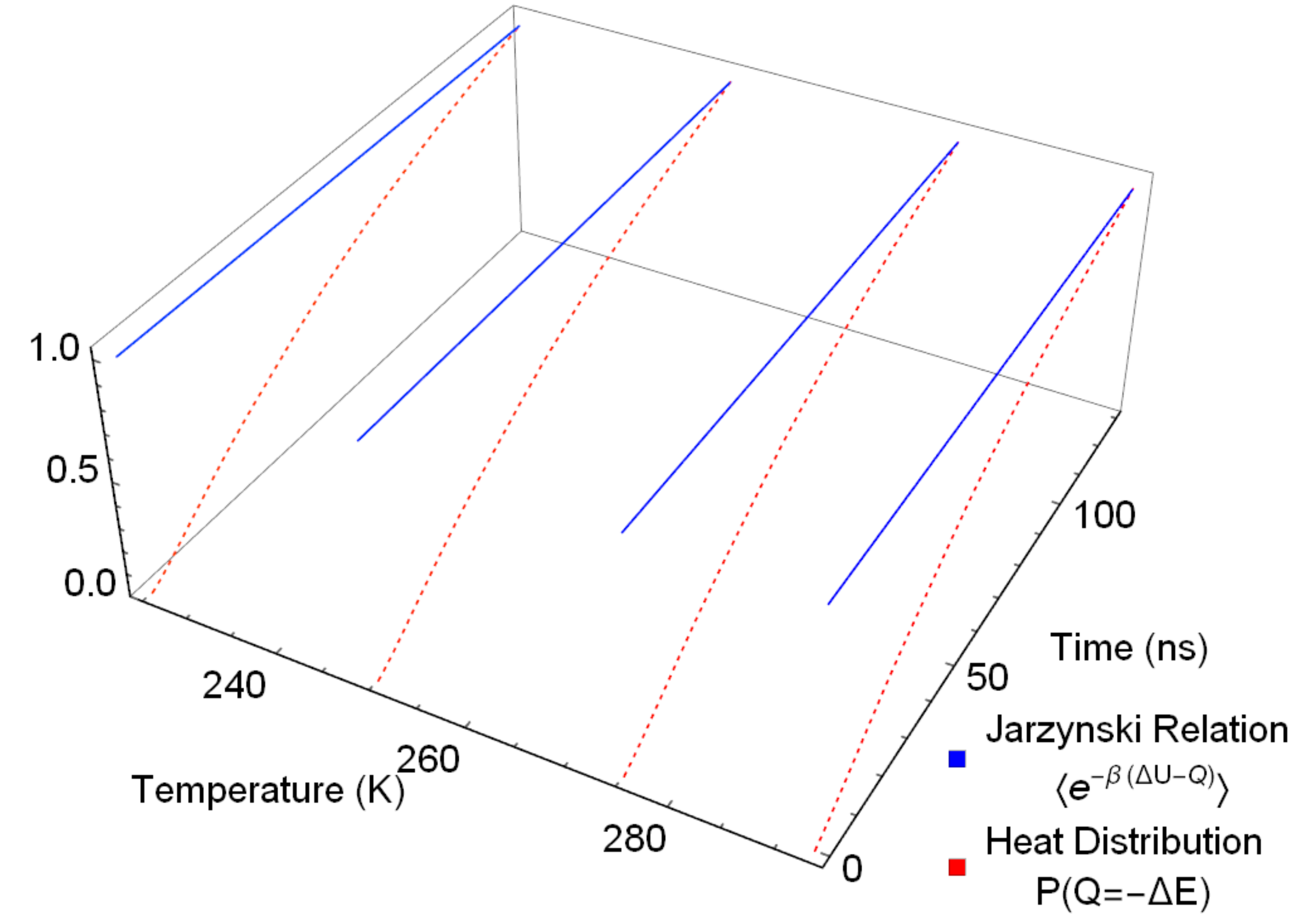}
\caption{\label{fig:dataplot}\textbf{Heat distribution and the Jarzynski equality} for an organic molecule undergoing spontaneous emission at a range of different temperatures. The solid curve is the result of overlaying the theoretical prediction of $\expt{\mathrm{e}^{-\beta \left(\Delta U - Q\right)}}=1$ with the experimentally calculated value. This time evolution is interpreted using Eq.~\ref{eq:heatdist}. The two results have such close agreement it is not possible to distinguish between them, demonstrating that the Jarzynski equality holds for all times during the emission. The lower dashed curve represents part of the heat distribution, which has been calculated directly from the spectroscopic data.}
\end{centering}
\end{figure}

We examine the spectroscopic data from spontaneous emission of the organic molecule dibenzoterrylene (DBT) in an anthracene crystal. The molecular system is effectively described by two energy eigenstates that are delocalized across the DBT. The molecule is coupled to a low excitation heat bath, and the joint system number state basis evolves unitarily under the action of:
\begin{equation}
\label{eq:unitary}
U = \begin{pmatrix}
			1 & 0 & 0 & 0 \\
			0 & \mu & -\nu^* & 0 \\
			0 & \nu & \mu^* & 0 \\
			0 & 0 & 0 & 1 
		\end{pmatrix},
\end{equation}
where $|\mu|^2 + |\nu|^2 = 1$. This coupling can be viewed as a partial swap between qubit and environment, parametrized by $\mu, \nu$. It has the effect of transferring $\Delta E = E_2-E_1$ ($E_i$ is the energy of the $i^\mathrm{th}$ level) between the molecule and the environment, with a probability dependent on the coupling strength. Within this experiment all of the photons are emitted along the zero phonon line, and as such we do not observe the effects of transitions within the vibrational modes. If we perform periodic measurements on the output field we can perform a partial measurement on the state of the molecule, and measure the heat flux from the molecule to the environment. This produces trajectories of the joint system which are shown in Fig. \ref{fig:trajplot}. We convert the time resolved trace of the photon statistics into the heat distribution which we find to be consistent with: 
\begin{eqnarray}
\label{eq:heatdist}
P(Q=\Delta E) &=& \alpha + \beta|\mu|^{2n} \nonumber \\
P(Q=-\Delta E) &=& \beta(1-|\mu|^{2n}),
\end{eqnarray}
where $\alpha, \beta$ are the initial occupation probabilities of each level. The heat distribution is plotted in Fig. \ref{fig:dataplot}, for two different laser intensities.

By examining the heat distribution given above we can directly evaluate the Jarzynski equality. In particular we find that the system satisfies $\expt{\mathrm{e}^{-\beta (\Delta U-Q)}}=1$ for all time points during the period of spontaneous emission. This demonstrates that the decay of organic molecules from an excited to the ground state is a thermodynamic process, even when it is considered at the level of single photon emission events.

We can utilize the Jarzynski equality to detect shifts in the energy spectra of the molecule. If we consider taking the energy levels from:
\begin{eqnarray}
\label{eq:energyshifts}
E_1 &\rightarrow& E_1 + \delta \nonumber \\
E_2 &\rightarrow& E_2 + \delta + \epsilon,
\end{eqnarray}
where $\delta$ is a common shift of the energies and $\epsilon$ is the difference in shifts between energy levels 1 and 2. This operation produces a change in the equilibrium free energy ($F=-k_B T\ln Z$) of the molecule: 
\begin{eqnarray}
\label{eq:deltaf}
\Delta F &=& k_\mathrm{B}T\ln\left(\frac{Z_f}{Z_i}\right) \nonumber \\
				 &=& -\delta + k_\mathrm{B}T\ln\left(\frac{1+\mathrm{e}^{-\beta \Delta E^{'}}}{1+\mathrm{e}^{-\beta \Delta E}}\right).
\end{eqnarray}
Here we have defined $\Delta E = E_2-E_1$ and $\Delta E^{'} = \Delta E + \epsilon$. By observing the fluorescence emitted by the excited molecules we can determine the value of both of these quantities. Combining this relation with the Jarzynski equality for heat we can determine the value of $\delta$. This would allow for probing the strengths of local field effects with high sensitivity.

\section{Discussion}


Recently there has been substantial interest in developing systems which can be used to test quantum thermodynamics, \cite{Batalhao14, An15}. Here we have demonstrated that the photon emission from organic molecules satisfies one of the fundamental relationships of the field. This is true for the single photon emission from the molecules, at all points during the relaxation time. We believe that this provides an important platform for the study of quantum thermodynamics in real systems. Organic spectroscopy is a well developed and established field of physics, and here we demonstrate that it possible to tap into these resources to investigate fundamental questions in quantum thermodynamics.

In the initial set up of this investigation we focused on emissions along the zero phonon line. One major advantage of investigating quantum thermodynamics by utilizing fluorescence from organic molecules is that it is experimentally straightforward to relax this assumption and probe how the situation changes when we incorporate the phononic transitions.  This would have the effect of creating an extra heat bath for energy transferred into the molecule to dissipate into, and would also provide for a much larger number of potential trajectories to be taken. With single molecule spectroscopy it is possible to have sufficient resolution of the photons emitted to determine how much energy is being ``lost'' to the phonon modes. We would then be able to determine if the Jarzynski equality is still satisfied in this open system.

In this analysis of the molecule-heat bath system we have modelled it as initially being in a mixed state. We believe this initial condition to be reasonable as the lifetime of any entanglement between the molecule and environment is very short lived. If instead the molecule were to be left in a superposition of energy eigenstates (e.g. $\ket{\psi}=\ket{g0}+\ket{e0}$) after the work is done on it then in the current setup there is no difference in the measured heat statistics relative to the corresponding de-phased mixed state. The reason this occurs is due to the repeated suppression of the coherence in the heat bath via measurement in the photon number basis and discarding the measurement outcomes. However as the initial state before the spontaneous emission is very close to a pure product state between the molecule and the heat bath, and the Hamiltonian time evolution is unitary, the total state before measurements is approximately pure, and there will exist a basis in which measurements would not destroy coherence. As is well known, see e.g.~\cite{Korzekwa16}, coherence implies extra extractable work so it would be valuable to identify how to implement this alternative basis measurement experimentally. 


\section{Methods}

Consider the two level molecule. We assume it is initially in a thermal state at the same temperature as the environment. In this set-up and for the modes of concern this means both the molecule and the environment are essentially in the ground state with a very small probability of a thermal excitation. The system is then pulsed by a laser which causes a swap in the occupation probabilities of the joint state. (The laser is non-resonant but there is then near-instantaneous vibrational de-excitation to the excited level of interest-see Fig. 2.) 
 Now the joint system undergoes evolution under its own Hamiltonian which has the effect of inducing a partial swap, parametrized by $|\mu|^2$, see Eq.~\ref{eq:unitary}. This unitary arises naturally from the Jaynes-Cummings Hamiltonian which describes the coupling between the environment and the molecule \cite{Scully97}. This corresponds to the emission of a photon to the environment as the molecule is de-excited.

Framing this in the language of quantum jumps, such that we can evaluate the Jarzynski equality \`{a} la \cite{Quan08}, produces the trajectories in Table \ref{tab:trajtable}. These trajectories are defined by considering an infinitesimal time step during which one of two events happens - either the system evolves under the joint Hamiltonian or we perform a projective measurement on the environment. The sequence of pure states produces a trajectory. 
\begin{table}[h!]
\begin{centering}
\begin{tabular}{|c|c|c|c|c|c|}
\hline
\begin{tabular}[c]{@{}c@{}}\end{tabular} & \begin{tabular}[c]{@{}c@{}}\textbf{Traj}\end{tabular} & \begin{tabular}[c]{@{}c@{}}\textbf{$W$}\end{tabular} & \begin{tabular}[c]{@{}c@{}}\textbf{$Q$}\end{tabular} & \begin{tabular}[c]{@{}c@{}}\textbf{$\Delta U$}\end{tabular} &
\begin{tabular}[c]{@{}c@{}}\textbf{Prob}\end{tabular}\\ \hline
1                   & $\ket{g}\rightarrow\ket{e}\rightarrow\ket{g}$              & $\Delta E$        & $-\Delta E$       & 0          & $P(\ket{g})P_{eg}$\\ \hline
2                   & $\ket{g}\rightarrow\ket{e}\rightarrow\ket{e}$  & $\Delta E$        & 0               & $\Delta E$  & $P(\ket{g})P_{ee}$\\ \hline
3                   & $\ket{e}\rightarrow\ket{g}\rightarrow\ket{e}$ & $-\Delta E$       & 0               & $-\Delta E$ &$P(\ket{e})P_{ge}$\\ \hline
4                   & $\ket{e}\rightarrow\ket{g}\rightarrow\ket{g}$            & $-\Delta E$       & $\Delta E$        & 0         &$P(\ket{e})P_{gg}$\\ \hline
\end{tabular}
\caption{
\label{tab:trajtable}
\textbf{Possible trajectories} of the local state of the molecule. }
\end{centering}
\end{table}

In this framework we consider the system to always be occupying a definite state and determine what possible evolutions each state can take. Each of these evolutions constitutes a possible trajectory the system can undertake, which have associated values of internal energy change, work and heat.

Having identified what the possible trajectories are, we must now determine which correspond to the experimental data we obtain. As we are detecting photon emission from the molecules using an avalanche photodiode (APD), in principle we detect trajectories 1, 3 and 4. However due to the electronics of the APD there is a delay time in receiving the signal from the photons \cite{bh15}. Trajectories 3 and 4 occur on a very fast time scale due to the stimulating effect of the laser causing the transition and so are unobserved. As such, we only observe clicks in the detector due to traj. 1. The probabilities of the remaining trajectories turn out, as described below, to be determined uniquely by the initial state being assumed to be thermal together with demanding that the probabilities of all the trajectories sum to 1.

The time resolution is very high and the probability of receiving more than one photon in a given time-bin is negligible here, even taking into account the number of molecules scattered in the illuminated part of the crystal. Thus one click means one photon.

\begin{figure}[h!]
\begin{centering}
\includegraphics[width=\columnwidth]{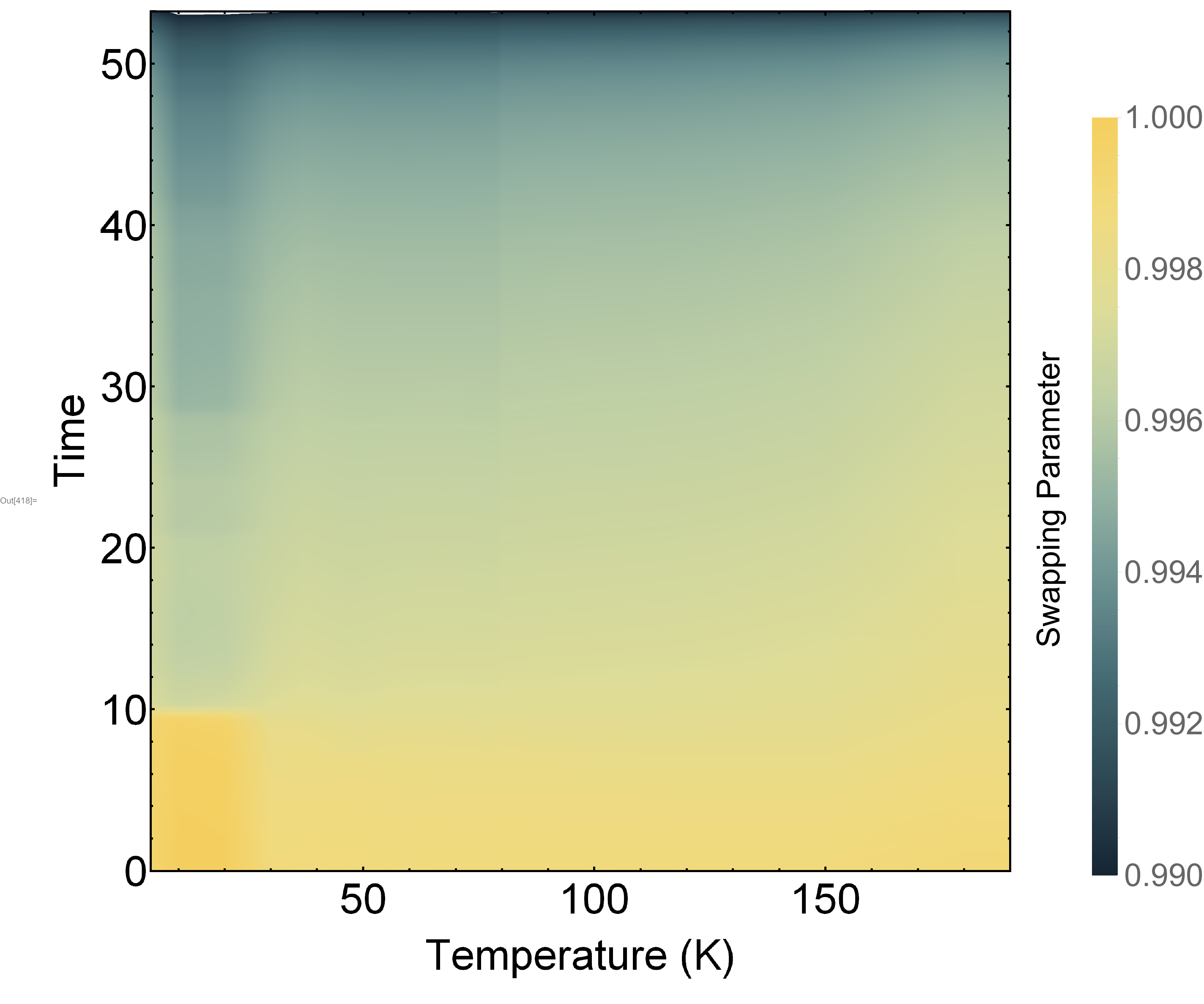}
\caption{\label{fig:swapplot}\textbf{The swapping parameter} that controls the interaction with environment is plotted as a function of temperature and the time. It can be seen that the swapping parameter is very stable for the majority of time during the experiment, indicating that the model we employ to interpret the data is sound, see Eq.~\ref{eq:unitary}. We utilize this in connecting the experimental data to the quantum jumps model that enables us to evaluate the Jarzynski equality.}
\end{centering}
\end{figure}

Having established the concurrence between clicks and photons, we now treat the environment as a low excitation heat bath. The emitted photons are then simply the heat flux into the bath, and by relating the cumulative probability of emission to the heat distribution, Eq.~\ref{eq:heatdist}, we can determine the values of $\alpha,\:\beta$ and $|\mu|^2$, see Fig. \ref{fig:swapplot}. This information corresponds to trajectories 1 and 2, and by employing the fact that $P(\mathrm{Traj}\:3)+P(\mathrm{Traj}\:4)=P(\ket{e})$ we can explicitly calculate the Jarzynski equality. This equality is plotted in Fig.~\ref{fig:dataplot}.

\section*{Conclusion}
We conclude that this fluorescence experiment amounts to a scenario where Jarzynski's equality holds. As further evidence that this experimental set-up is promising for probing quantum thermodynamics we note/reiterate the following: (i) molecular spectroscopy is a powerful and well-established technique, (ii) the set-up maps naturally to the quantum jumps model for work and heat (used e.g.\ in Crook's proof of Quantum Jarzynski \cite{Crooks99}). A natural and exciting next step in this program is to prepare two entangled molecules to implement the quantum Maxwell's demon of \cite{delRio11}. This would build on the important first steps we have taken with this work, and reinforce the applicability of quantum thermodynamics to spectroscopy. A further area of research would be exploring the technological applications of utilizing these organic molecules to probe external fields by employing the relationship between spectroscopy and the Jarzynski equality.

\section*{Acknowledgements}
We are grateful for comments from John Goold and Mark Mitchison. The authors thank the Oxford Martin School, Wolfson College and the University of Oxford, the Leverhulme Trust (UK), the John Templeton Foundation, the EU Collaborative Project TherMiQ (Grant Agreement 618074), the COST Action MP1209 and the EPSRC (UK). This research is also supported by the National Research Foundation, Prime Minister's Office, Singapore, under its Competitive Research Programme (CRP Award No. NRF- CRP14-2014-02) and administered by Centre for Quantum Technologies, National University of Singapore. 

\bibliographystyle{unsrt}

\end{document}